\documentclass[twocolumn,english,prl,showpacs]{revtex4}
\usepackage{graphicx}
\usepackage{mathrsfs}

\begin{document}
\title{Phase separation in Bose-Fermi-Fermi Mixtures as a probe of
  Fermi superfluidity} \author{S. G. Bhongale} \affiliation{Department
  of Physics \& Astronomy, and Rice Quantum Institute, Rice
  University, Houston, TX 77005, USA}\email{bhongale@rice.edu}
\author{H. Pu}\affiliation{Department of Physics \& Astronomy, and
  Rice Quantum Institute, Rice University, Houston, TX 77005, USA}
\begin{abstract}
We study the phase diagram of a mixture of Bose-Einstein
condensate and a two-component Fermi gas. In particular, we
identify the regime where the homogeneous system becomes unstable
against phase separation. We show that, under proper conditions, the
phase separation phenomenon can be exploited as a robust probe of
Fermi superfluid.

\end{abstract}

\pacs{67.85.Pq,03.75.Nt,64.70.Tg,67.25.D-} \date{\today} \maketitle

Mixtures of superfluids open up possibilities of studying interacting
macroscopic quantum systems.  The attempt of such studies started
decades ago in the system of $^3$He-$^4$He mixtures. However, the
transition into the superfluid phase of $^3$He atom in these mixtures
occurs at an extremely low temperature and has never been reached in
experiment. Realization of superfluids in atomic quantum gases makes
such studies possible for the first time.

The atomic analogy of a superfluid $^3$He-$^4$He system is a mixture
of Bose-Einstein condensate (BEC) and superfluid Fermi gas (i.e., a
two-component Fermi gas with attractive interaction). In this paper,
we explore the rich phase diagram of this system at zero
temperature. By studying the free-energy of the homogeneous mixture,
we identify the regime where the homogeneous mixture becomes
unstable against phase separation. Phase separation is quite a
generic phenomenon occurring in trapped atomic mixtures, originating
from the interplay between the interactions and the spatial
variation induced by the trap, thus allowing different regions of
the trap to favor fundamentally different phases
\cite{timmermans,fermiseparation}. In addition, we demonstrate a
novel application of the phase separation phenomenon- the detection
of Bardeen-Cooper-Schrieffer (BCS) superfluidity within the Fermi
gas \cite{bcstheory,bcsexpts}. Such a detection proposal is
motivated by the need for an efficient superfluidity detection probe
for understanding pairing within unbalanced Fermi mixtures, a
scenario that has recently attracted a lot of attention.

\begin{figure}[t]
  \includegraphics[scale=.30]{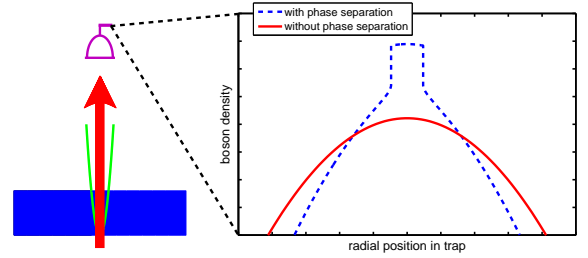}
\caption{(Color online) Schematic of the proposed BEC probe.  Bosons
  are confined in the tight trap and are made to interact with the
  Fermi atoms which are unconfined.  Phase separation can be easily
  detected by measuring the BEC density profile.}
\label{cartoon}
\end{figure}

The probing concept can be simply understood if we envision a BEC
localized to a small region within the Fermi medium. We assume that
the Fermi gas is unconfined which is a valid approximation if the
spatial extension of the Fermi cloud is much broader than that of
the BEC.  In the phase separation regime, the BEC may exist in a
pure form as an isolated bubble surrounded by the Bose-Fermi
mixture. We show that the formation of such a BEC bubble may be made
sensitive to the superfluid property of the fermionic medium. Thus,
the onset of BCS superfluidity is signalled by the formation of
small BEC bubbles which can be readily detected via a density
measurement. Essentially, bosons serve as a matter-wave probe of
Fermi superfluid. This idea is illustrated schematically in
Fig.~\ref{cartoon}.

To begin with, we first determine the zero-temperature projected phase
diagram of a Bose-Fermi-Fermi mixture comprising of bosons of one
species and equal population spin-up and -down fermions of
another. Here, by projected we mean a $2$-dimensional slice of the
$d$-dimensional parameter space by fixing the $d-2$ independent
parameters. In the absence of inter-fermion interaction, this system
can be identically mapped onto a Bose-Fermi mixture and has been
previously studied quite extensively \cite{bosefermiphase}.  However,
the interacting case remains much less explored \cite{bfs} and is of
focal importance to this work. Here, we therefore assume attractive
$s$-wave interaction between fermions of unlike spins. Furthermore, we
assume Bose-Bose and Bose-Fermi interactions to be repulsive.  It is
convenient to treat all interactions via a pseudo-potential modeled by
$v_{\alpha}({\bf r}-{\bf x})=\lambda_{\alpha}\delta({\bf r}-{\bf x})$
which is valid for dilute systems, where we indicate different types
of interacting atoms by the index $\alpha \in
\{BB,B\uparrow,B\downarrow,\uparrow\downarrow \}$. In this work, we
make two non-essential simplifications. First, we assume that the
Bose-Fermi interaction is spin-independent, i.e.,
$\lambda_{B\uparrow}=\lambda_{B \downarrow} = \lambda_{BF}$.  Second,
we consider a single spatial dimension. We remark that our theoretical
framework is general and the above two restrictions can be
straightforwardly removed.

The homogeneous system under study is characterized by a total of
5 independent parameters: the 3 interaction strengths
$(\lambda_{BB}, \lambda_{BF}, \lambda_{\uparrow \downarrow})$ and
two densities $(\varrho_B, \varrho_F)$ where $\varrho_\uparrow
=\varrho_\downarrow=\varrho_F/2$. Note that the choice is not
unique. For instance, instead of the densities, we may choose
chemical potentials $(\mu_B, \mu_F)$ to be independent parameters.
It is instructive to first consider the much simpler case without
Fermi-Fermi interaction, or $\lambda_{\uparrow\downarrow}=0$. As
mentioned earlier, this case, for 3D confinement, has been studied
by several authors \cite{bosefermiphase}. Similar analysis can be
performed for the 1D case, resulting in the instability criterion
given by $\varrho_F<4\lambda_{BF}^2/(\pi^2\lambda_{BB})$ --- when
this inequality is satisfied, the homogeneous mixture is unstable
and tends to phase separate. Thus the BEC density profile in the
two regimes (phase separated or not) may be made quite distinct as
depicted in the schematic of Fig.~\ref{cartoon}. However, we
emphasize that, since the interaction is density-density, there is
no direct connection of the fermionic pairing gap (in the case of
interacting fermions) to the {\it phase-separation} phenomenon.
Remarkably, as we shall show, the phase separation {\em can} be
made sensitive to the fermionic pairing, underlying the basis of
our proposed BEC probe of Fermi superfluidity.

We begin by writing the free energy functional of the homogeneous
mixture in the form,
\begin{widetext}
\begin{equation}
{\bf F}[\varrho_B,\Delta]=\frac{\lambda_B}
{2}\varrho_B^2-\mu_B\varrho_B(r)+\frac{4}{2\pi}\int_0^{\infty}\frac{\varepsilon_k
  -\tilde{\mu}_F}{2}\left[1-\frac{\varepsilon_k-\tilde{\mu}_F}{\Lambda_k}\right]
d k+\lambda_{\uparrow \downarrow} \Delta^2
\left[\frac{1}{2\pi}\int_0^{\infty}\frac{1}{\Lambda_k}d
  k\right]^2, \label{free}
\end{equation}
\end{widetext}
where we have defined $\varepsilon_k=\hbar^2k^2/2m$,
$\tilde{\mu}_F=\mu_F-\lambda_{BF}\varrho_B$, and the fermionic
quasi-particle energies are given by
$\Lambda_k=\sqrt{(\varepsilon_k-\tilde{\mu}_F)^2+\Delta^2}$ by
neglecting the Hartree contribution since it only leads to a
constant energy shift. Also, anticipating the role of Fermi
superfluid, we have written the free energy as a function of boson
density $\varrho_B$ and BCS pairing gap $\Delta$. We will then
construct the projected phase diagram in the $\varrho_B$-$\Delta$
parameter space while fixing the values of three more independent
variables to be discussed below.

The thermodynamic ground state is given by the minimum of the the free
energy and therefore corresponds to the necessary first derivative
conditions: $\partial {\bf F}/\partial \Delta=0$ and $\partial {\bf
  F}/\partial \varrho_B=0$.  The first of these conditions essentially
reproduces the gap equation in the form $ (-
\lambda_{\uparrow\downarrow}/2\pi)\int_0^\infty (1/\Lambda_k) dk=1 $,
while the second fixes the number through the modified bosonic
Thomas-Fermi equation
$\lambda_B\varrho_B-\mu_B+\lambda_{BF}\varrho_F\nonumber=0 $.
However, local minimum is guaranteed only if the Hessian matrix ${\bf
  M}$, constructed from the second derivatives of ${\bf F}$, is
positive definite, or the following conditions are satisfied:
\begin{equation}
{\bf
  M}_{\varrho_B\varrho_B} =\frac{\partial^2{\bf F}}{\partial
  \varrho_B^2} >0 \,,\;\;\;{\rm Det}[{\bf M}] >0 \,. \label{saddlecondition}
\end{equation}
When condition (\ref{saddlecondition}) is violated, the system will
necessarily phase separate \cite{note}. Using this simple criteria,
one can map the whole phase space of the homogeneous mixture. However,
this is an extremely laborious task, given that the total phase space
is huge, represented by the set of five independent parameters as we
mentioned earlier. We make a judicious choice and use the following
set: $\{\lambda_{BB}, \lambda_{BF}, \Delta, \varrho_B, \mu_F\}$, the
motivation behind which will be clear as we proceed further. Other
parameters, such as $\lambda_{\uparrow \downarrow}$, $\varrho_F$ and
$\mu_B$, must be calculated self-consistently using the gap equation
and the bosonic Thomas-Fermi equation discussed earlier, together with
the fermionic number equation $ \varrho_F=(2/\pi)\int_0^{\infty}
[1-(\varepsilon_k-\tilde{\mu}_F)/\Lambda_k ]dk$.

We remind the reader that our goal here is not to give a complete
description of the whole phase space but focus our study on a small
region that is physically meaningful in view of current experimental
setups and illustrate a novel probing techniques based on the phase
separation phenomenon. This we do by first picking reasonable values
of $\lambda_{BB}$, $\lambda_{BF}$, and $\mu_F$. We then determine
the stable/unstable regions in the $\varrho_B$-$\Delta$ space via
condition (\ref{saddlecondition}). For this we need to study the
properties of the Hessian matrix ${\bf M}$. States that satisfy
condition (\ref{saddlecondition}) are only guaranteed to be a local
minimum of the free energy. To determine whether the state is the
ground state of the system, we need to compare the free energies of
different homogeneous phases which include the pure BEC and the pure
Fermi phase, in addition to their mixture.


Following the above procedure, we have determined the phase space
for various values of the fixed parameters. We find that this system
exhibits a very rich phase diagram. However given the lack of space,
we restrict ourselves to pointing out some general features that are
relevant to this proposal and direct the reader to an upcoming
publication for more details.

\begin{figure*}[t]
   \includegraphics[scale=.56]{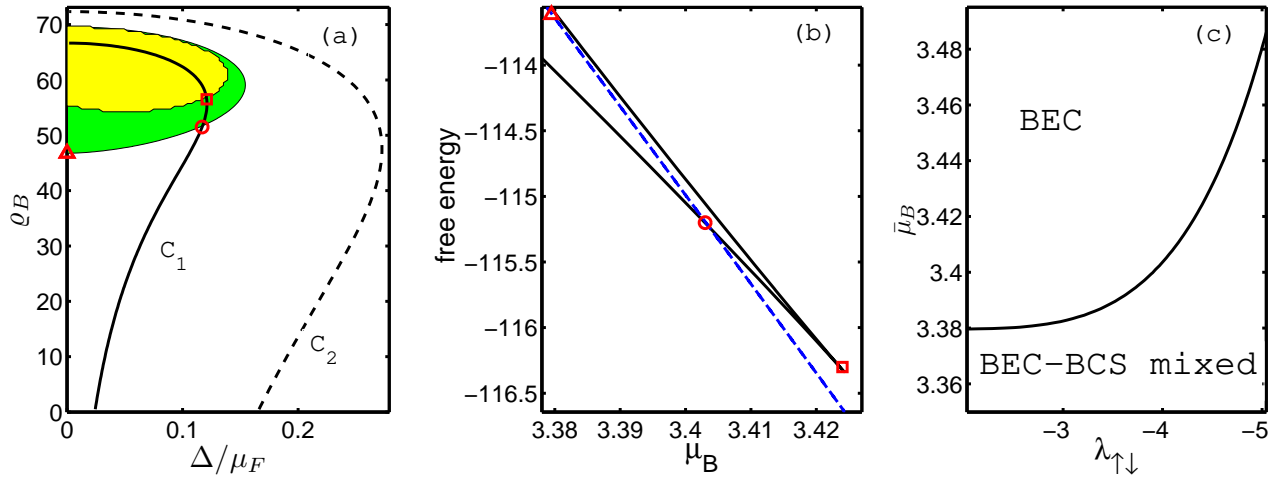}
\caption{
(Color online) (a) Projected phase space of the
  Bose-Fermi-Fermi mixture. Here, as well as in other figures, all
  quantities are scaled in units of the probe, represented by an
  harmonic oscillator with frequency $\omega_0$ and length
  $\ell_0$. Fixed parameters are $\lambda_{BF}=0.44$,
  $\lambda_{BB}=0.05$, and $\mu_F=27.32$. The yellow (green) area
  indicates region where the Bose-Fermi mixture is dynamically
  (mechanically) unstable. The contour, ${\bf C_1}$ and ${\bf C_2}$,
  corresponding to $\lambda_{\uparrow\downarrow}=-4$ and $-6$
  respectively. (b) Comparison of free energy of the mixture (solid
  curve) along the contour ${\bf C}_1$ and the free energy of the pure
  BEC (dashed curve). The free energy the pure Fermi gas is much
  higher and not shown. The points shown by red symbols correspond to
  those in (a). (c)Critical boson chemical potential,
  $\bar{\mu}_B$ as a function of the inter-Fermi interaction
  separating the pure BEC and mixed Bose-Fermi phase. Within LDA,
  moving vertically upwards along a line parallel to the $y$ axis
  implies moving from the edge towards the center of the trap.}
\label{stableregion}
\end{figure*}

The projected phase diagram of a homogeneous mixture for a
particular set of $\{\lambda_{BF}, \lambda_{BB},\mu_F\}$ is shown in
Fig.~\ref{stableregion}(a). We note the following: (1) The unstable
region correspond to a partial ellipse cut by the $\varrho_B$ axis
(shown by the shaded region in Fig.~\ref{stableregion}(a)). (2) Within
this region, the phase space can be further divided into a
dynamically unstable region (the yellow/brighter shaded) at which
the free energy landscape shows a saddle point and a dynamically
stable region (the green/darker shaded) at which the free energy
landscape shows a local but not a global minimum. (3) The position
and the extent of these regions depends on the values of the fixed
parameters. Immediately we notice that, in a typical experimental
setup with a given value of $\lambda_{\uparrow\downarrow}$, only a
small part of the phase space comprising of points on the fixed
$\lambda_{\uparrow\downarrow}$ contour, for example ${\bf C}_1$ or
${\bf C}_2$ shown in Fig.~\ref{stableregion}(a) is physically
accessible. The vertical axis of $\Delta=0$ corresponds to a mixture
of BEC with a non-interacting Fermi gas ($\lambda_{\uparrow
\downarrow}=0$). The significance of the particular parameter set
chosen earlier is clear since we can now directly obtain the Bose
density profile in the probe by mapping the boson chemical potential
along this contour, $\mu_B[{\bf C}_{i}]$ on to the spatial
coordinate in the probe via the local density approximation (LDA)
using $\mu_B[{\bf C}_i]=\mu_B(r) \equiv \mu_B-V(r[{\bf C}_i])$ where
$V(r)$ is the probe trapping potential for BEC. In
Fig.~\ref{stableregion}(b) we plot the free energy of the mixture as a
function of the Bose chemical potential $\mu_B[{\bf
    C}_1]$. The same plot also show the
free energy for a pure BEC. Now we can easily identify the following
special points in Fig.~\ref{stableregion}(a): the point marked by the
circle(triangle) as the point where the free energy of the BEC goes
below that of the mixture and thus represents a first order phase
transition in the presence(absence) of pairing, and the point marked
by square as the point where $\mu_B$ reaches its maximum value along
the contour ${\bf C}_1$.



Now we are in the position to discuss how we can take advantage of the
phase diagram to detect Fermi superfluidity using BEC as a probe.  To
put this idea in context, we note that there is a new avenue in cold
atom research that cast BEC as tools for quantum measurement. A good
example is the experiment by Kr\"{u}ger {\em et al.}
\cite{schmiedmayer} where the Thomas-Fermi character of the BEC
density profile is exploited to measure the surface potential energy
landscape with exquisite accuracy. Also, very recently, Bhongale and
Timmermans have proposed a high sensitivity force measurement by
exploiting phase separated two-BEC mixture \cite{bhongaleandtimm}. The
ideas in this letter are a next step in this direction.

\begin{figure}[t]
  \includegraphics[scale=.55]{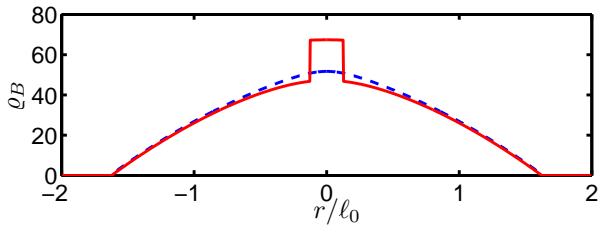}
\caption{(Color online) Boson density profile within LDA with (dashed)
  and without(solid) fermion pairing, for total number of bosons
  $N_B=101$, at $\lambda_{\uparrow\downarrow}=-4$. All other
  parameters are same as used in the previous plot.}
\label{densityplot}
\end{figure}

Since for our system the bosons are confined to the probe, we want a
situation such that, on phase separation, the system consists of a
pure BEC component near the center of the trap surrounded by a cloud
of either mixed bosons and fermions or pure fermions.  It is quite
intuitive that if $\mu_B(r)>
\bar{\mu}_B[\lambda_{\uparrow\downarrow}]$ (the latter being the boson
chemical potential along the stable boundary of the green region
represented by the circle in Figs.~\ref{stableregion}(a) and
\ref{stableregion}(b), a pure BEC bubble will phase separate out from
the mixture. However, in order to use this phenomenon to discern the
fermion superfluid property, it is important to connect the occurrence
of the BEC bubble with the disappearance/appearance of a nonzero
superfluid gap $\Delta$.  This is crucial since, as mentioned earlier,
phase separation can also occur in a mixture of BEC and a {\em normal}
Fermi gas due to Bose-Fermi repulsion. For this reason, in
Fig.~\ref{stableregion}(c), we plot the critical boson chemical
potential as a function of the inter-fermion interaction strength. We
see that there is a clear separation of regions corresponding to a
pure BEC phase above the curve and a mixed phase below. Thus if we
configure the probe such that the center chemical potential is very
close to and above the curve in Fig.~\ref{stableregion}(c) implying a
phase separated pure BEC component near the center of the trap, the
appearance of BCS superfluidity on increasing the attractive
inter-fermion interaction is immediately signalled by the disappearance
of the phase separated pure BEC bubble as shown in
Fig.~\ref{densityplot}.

In, practice, the sharp density jump will be smoothed out due to the
finite kinetic energy contribution, however, the absence of non-Thomas
Fermi variation in the Boson density profile can be easily
detected. Also, we claim that this probing method allows for a numeric
estimation of the gap, $\Delta$, by a systematic fitting technique if
the exact density profile is obtained via a numerical solution of the
coupled Bose-Fermi equations.

Moreover, increasing the attractive interaction beyond a certain value
(depending on the value of $\lambda_{BB}$ and $\lambda_{BF}$) results
in the absence of a phase separated regime as depicted by the contour
${\bf C}_2$ in Fig.~\ref{stableregion}(a). In fact, this can be
considered as a very strong indication of the pairing phenomenon, the
limiting case of which is a homogeneous mixture of molecular (M) and
atomic BEC on the repulsive side of the Feshbach resonance, known to
be completely stable if $\lambda_{BM}<\sqrt{\lambda_{BB}\lambda_{MM}}$
\cite{timmermans}.  However, this involves calculating these
additional interaction strength, which will be dealt with in a future
article.

In conclusion we have studied the phase diagram of a
Bose-Fermi-Fermi mixture and proposed a probe for detecting BCS type
superfluidity within a quasi-1D two-component Fermi gas by
configuring the system such that thermodynamic instability resulting
in phase-separation is sensitive to the interaction between the
fermions and hence the BCS pairing. The probe consists of a BEC
confined to a relatively tight trap. We have shown that by properly
tuning probe parameters, the density profile of the BEC provides a
robust signal of the fermion pairing. The probe idea may be easily
extended to 3D systems by identifying appropriate phase separation
regime. The expressions for the free energy in Eq.~(\ref{free}) and
the corresponding Hessian matrix are still valid in 3D with the
understanding that the integrals involved are also 3D. One important
difference between 3D and 1D is that, in the former, proper
renormalization procedures must be taken to remove the ultraviolet
divergence in gap equation associated with the contact interaction.

We are aware of a related recent proposal for probing BCS type
superfluidity using a overlapping BEC. However there, the pairing
signal is related to the damping of the BEC acoustic phonons
\cite{gaudio}. The strength of our proposal lies in the fact that
the signature of pairing is reflected in the BEC density profile,
a quantity that can be easily measured in experiment. The proposed
probing scheme possesses another important advantage: it probes
the local value of $\Delta$. This is crucial in trapped experiment
where the gap varies in space due to the trap-induced
inhomogeneity, and in situation where the gap has intrinsically
nontrivial spatial dependence. The latter arises, for instance, in
the case of FFLO superfluid state in a population imbalanced Fermi
system where the gap varies sinusoidally in space, which may
result in density oscillations in BEC probe. Another natural
extension is to study the phase diagram of the Bose-Fermi-Fermi
mixture where the Fermi-Fermi interaction is tuned across a
Feshbach resonance. How the presence of the bosons affect and
probe the BEC-BCS crossover will be an interesting problem to
study.

We thank Randy Hulet for useful discussions. This work is
supported by the W. M. Keck Program in Quantum Materials at Rice
University, NSF and the Robert A. Welch Foundation (Grant No.
C-1669).


\begin{thebibliography}{99}
\bibitem{note}An unstable homogeneous state may also tend to
collapse. This possibility, however, is excluded by our choice of
repulsive Bose-Bose and Bose-Fermi interactions.
\end{thebibliography}
\end{document}